# CCAT-prime: Science with an Ultra-widefield Submillimeter Observatory at Cerro Chajnantor

G.J. Stacey[1a], M. Aravena[b] K. Basu[c], N. Battaglia[a,d], B. Beringue[e], F. Bertoldi[c], J. R. Bond[f], P. Breysse[f], R. Bustos[g], S. Chapman[h,i], D. T. Chung[j], N. Cothard[k], J. Erler[c], M. Fich[l], S. Foreman[f], P. Gallardo[m], R. Giovanelli[a], U.U. Graf[n], M.P. Haynes[a], R. Herrera-Camus[o], T.L. Herter[a], R. Hložek[p], D. Johnstone[i], L. Keating[f], B. Magnelli[c], D. Meerburg[e,q,r,s], J. Meyers[f], N. Murray[f], M. Niemack[m], T. Nikola[t], M. Nolta[f], S. C. Parshley[a], D. Riechers[a], P. Schilke[n], D. Scott[h], G. Stein[f], J. Stevens[m], J. Stutzki[n], E.M. Vavagiakis[m], M.P. Viero[j]

[a]Dept. of Astronomy, Cornell Univ., Ithaca NY, 14853 USA; [b]Núcleo de Astronomía, Facultad de Ingeniería y Ciencias, Univ. Diego Portales, Santiago 8370191, Chile; [c]Argelander Inst. for Astronomy, Rheinishche Friedrich-Wilhelms Univ. of Bonn, D-53121 Bonn, Germany; [d]Center for Computational Astrophysics, Simons Foundation, NYC NY, 10010 USA; [e]DAMTP, Centre for Mathematical Sciences, Wilberforce Road, Cambridge, CB3 0WA, UK; [f]Canadian Inst. for Theoretical Astrophysics, 60 St George St, Toronto ON M5S 3H8, Canada; [g]Facultad de Ingeniería, Univ. Católica de la Santísima Concepción, Alonso de Ribera 2850, Concepción, Chile; [h]Dept. of Physics and Astronomy, Univ. of British Columbia, Vancouver BC, Canada; [i]National Research Council Canada, Herzberg Astronomy & Astrophysics, Victoria BC, Canada; [j]Kavli Inst. for Particle Astrophysics and Cosmology & Physics Dept., Stanford Univ., Stanford CA 94305, USA; [k]Dept. of Applied and Engineering Physics, Cornell Univ., Ithaca NY, 14853 USA; [l]Dept. of Physics and Astronomy, Univ. of Waterloo, 200 University Avenue West Waterloo ON, N2L 3G1Canada; [m]Dept. of Physics, Cornell Univ., Ithaca NY, USA 14853; [n]I. Physical Inst., Univ. zu Köln, Zülpicher Str.77 50937 Köln, Germany; [o]Max-Planck-Institut fur extraterrestrische Physik, Giessenbachstr., 85748 Garching, Germany; [p]Dunlap Inst. for Astronomy & Astrophysics, Univ. of Toronto 5 - St. George Street, Toronto, Canada M5S 3H4; [q]Kavli Inst. for Cosmology, Madingley Road Cambridge CB3 0HA UK; [r]Kapteyn Astronomical Institute, Univ. of Groningen, P.O. Box 800, 9700 AV Groningen, The Netherlands; [s]Van Swinderen Inst. for Particle Physics and Gravity, Univ. of Groningen, Nijenborgh 4, 9747 AG Groningen, The Netherlands, [t]Center for Astrophysics and Planetary Science, Cornell Univ., Ithaca NY, 14853 USA

## ABSTRACT

We present the detailed science case, and brief descriptions of the telescope design, site, and first light instrument plans for a new ultra-wide field submillimeter observatory, CCAT-prime, that we are constructing at a 5600 m elevation site on Cerro Chajnantor in northern Chile. Our science goals are to study star and galaxy formation from the epoch of reionization to the present, investigate the growth of structure in the Universe, improve the precision of B-mode CMB measurements, and investigate the interstellar medium and star formation in the Galaxy and nearby galaxies through spectroscopic, polarimetric, and broadband surveys at wavelengths from 200 μm to 2 mm. These goals are realized with our two first light instruments, a large field-of-view (FoV) bolometer-based imager called Prime-Cam (that has both camera and an imaging spectrometer modules), and a multi-beam submillimeter heterodyne spectrometer, CHAI. CCAT-prime will have very high surface accuracy and very low system emissivity, so that combined with its wide FoV at the unsurpassed CCAT site our telescope/instrumentation combination is ideally suited to pursue this science. The CCAT-prime telescope is being designed and built by Vertex Antennentechnik GmbH. We expect to achieve first light in the spring of 2021.

---

[1] gjs12@cornell.edu; phone 1 (607) 255-5900

## 1. INTRODUCTION

CCAT-prime is a 6-m aperture submillimeter (submm) to millimeter (mm) wave telescope sited on a 5600 m elevation plateau about 40 m below the summit of Cerro Chajnantor in northern Chile, inside the Parque Astronómico de Atacama [1]. The telescope optics are based on the off-axis crossed-Dragone [2] design, which yields an extraordinarily wide FoV for astrophysical applications. The science requirement is very high surface accuracy (rms half wave-front-error < 10.7 μm, 7 μm (goal)), with no blockage in front of the receiver, and very small gaps between telescope panels, so as to fulfill our telescope emissivity requirement of less than 2.8% (goal < 1%). The high surface accuracy, low-emissivity telescope at the superb site will deliver superior surface brightness sensitivity in the mm to submm atmospheric windows. Combined with the wide FoV, CCAT-prime will have unrivaled mapping speed in these windows enabling the following science:

1. Through intensity mapping over tens of square degrees in spectral lines such as the redshifted 158 μm [CII] line, CCAT-prime will reveal the formation, growth, and three dimensional large scale clustering properties of the first star forming galaxies from redshifts of 9.3 (in the epoch of reionization) to redshifts of 3.3 (near the epoch of peak star formation), and those at lower redshifts in CO rotational lines.
2. Through multi-frequency measurements in the 100 to 860 GHz band of the Sunyaev-Zeldovich effect for more than 1000 galaxy clusters, CCAT-prime will measure the physical properties and spatial distribution of such clusters, placing constraints on fundamental physics including the nature of dark energy and the sum of the neutrino masses, and revealing the effects of active galactic nuclei-star formation feedback in clusters.
3. Through multi-frequency characterizations on the foreground dust polarization, CCAT-prime will greatly improve limits on primordial gravitational waves and constraints on inflationary models obtained through CMB polarization measurements.
4. Through multi-frequency photometric measurements of dusty sources of emission, CCAT-prime will trace the history of dusty star formation deeply into the epoch of galaxy formation more than 10 billion years ago.
5. Through high frequency spectrally resolved [CI] and CO line mapping, CCAT-prime will reveal the physical processes associated with star formation in various environments in the Milky Way, the Magellanic Clouds, and other nearby galaxies. We will also detect the flickering light of accretion disks enveloping protostellar envelopes through changes in their submm continuum brightness.

The first four science goals and the second half of goal 5 are pursued with our first-light camera Prime-Cam. Prime-Cam is based on seven independent "optics tubes" within a single cryostat, each tube addressing a 1° to 1.3° diameter FoV with bolometer arrays targeting specific science goals. Due to resource limitations we will initially populate three of these tubes. The central tube will be an 860 GHz broad-band camera that uses kinetic inductor detector (KID) technologies. When fully populated it will have 18,216 pixels with a beam size of ~14". The outer ring will contain a combination of multi-chroic (220, 270, 350 & 405 GHz) camera optics tubes and imaging spectrometer tubes, both based on TES bolometer arrays. The total TES bolometer counts are 8640 (camera) and 4320 (spectrometer) per tube. At first light, we plan at least one multi-chroic camera and one imaging spectrometer tube. We also have a first-light 64 pixel high-frequency heterodyne spectrometer, the CCAT Heterodyne Array Instrument (CHAI), which addresses science goal 5. The instrumentation is described in more detail in section 4 and in a companion publication [3].

Table 1 shows the mapping of science goals to instrument modules and survey areas. The required areas and integration times are based on detailed science simulations that fold in the wavelength coverage, sensitivity, and field-of-view of Prime-Cam (Figs. 1-5). The following sections describe the survey science, the site properties, the telescope design, and the first light, and future instrument concepts. We end with a brief discussion of our timeline to first light.

## 2. SCIENCE WITH THE CCAT-PRIME FACILITY

### 2.1 Science with the first light instrument Prime-Cam: the high redshift Universe

#### 2.1.1 What are the physical properties of the galaxies that reionize the Universe?

The Universe began as an expansion of spacetime from an initially hot and dense phase, the Big Bang. With expansion, density and temperature dropped so that at about 380,000 years after the origin (redshift, $z \approx 1100$) the gas/photon temperature had dropped to $T \approx 3000$ K, at which time free electrons could combine with protons forming neutral hydrogen. This epoch is therefore termed the "recombination epoch". At this time, the source of photon opacity – Thompson scattering off free electrons – therefore essentially vanished, and the photons and matter were thermally disconnected. The near perfect blackbody radiation that was released is measured today as the Cosmic Microwave Background (CMB), with a blackbody temperature of 2.73 K.

| Science Case[a] | Array Type | Frequencies[b] (GHz) | Spatial Resolution (arcsec) | Detectors Type | Detectors Number | Survey Areas[c] (deg$^2$) Pilot | Survey Areas[c] (deg$^2$) Full |
|---|---|---|---|---|---|---|---|
| SZ, CMB, SFH | polarimetry photometry | 220, 270, 350, 405 | 53, 46, 39, 37 | TES | 8640 | 100 | 12,000[d] |
| EoR, SZ | spectroscopy RP ~ 280 | 222, 270, 350, 405 | 53, 46, 39, 37 | TES | 4320 | 4 | 16 |
| SFH | photometry | 860 | 14 | KID | 18,216 | 100+4 | 12,000+16 |
| GEco | photometry spectroscopy RP ~$10^3$-5×$10^6$ | 455-495 800-820 | 26 15 | SIS SIS | 8 × 8 8 × 8 | 50 | 200 |

Table 1. How Science Cases Map onto the Detector Array Types and Planned Surveys

[a]Science cases: SZ: Clusters, SZ, and cosmology; CMB: CMB polarization; SFH: Star Formation History, the evolution of dusty star forming galaxies; EoR: intensity mapping of reionization; GEco: Galactic Ecology
[b]90- and 150-GHz data from Advanced ACTPol and Simons Observatory will be available by agreement.
[c]Areas for initial (400 hour) and multi-year (4000 hour) surveys; 860 GHz optics tube will be used with both survey types.
[d]Over 400 deg$^2$ of the survey will be observed down to the source confusion limit for the SFH survey.

The hydrogen in today's intergalactic medium is almost totally ionized, so it must have been reionized by early starlight or accretion luminosity onto massive black holes in AGN at early times. Several lines of research have established that the epoch of reionization (EoR) was a gradual process extending from approximately 200 to 900 million years after the Big Bang ($17 \geq z \geq 6$). This is the first and least explored epoch in which stars and galaxies began to govern the shape and overall properties of the Universe [4]. Studies of individual galaxies with HST and ALMA hint that cosmic reionization may be driven by the first galaxies that form in the Universe, but the overall properties and distribution of these sources of reionization are still poorly understood [5]. This is because the main sources of reionization are very numerous, but intrinsically extremely faint. Indeed, as shown by the recent claimed detection of unexpectedly strong and high redshift HI 21-cm absorption signal at z =17 – near the onset of reionization [6] – our understanding is still incomplete, so that direct measurements of all phases of reionization are critical. To make significant progress, large-scale three-dimensional surveys of the Universe during the EoR are of fundamental importance. Such surveys become possible by employing new observing strategies, in particular spectral line intensity mapping [7]. Line-intensity mapping is a technique that measures the spatial fluctuations of signals due to large-scale structure at low spatial resolution, providing three-dimensional spatial information of the sources of emission (or absorption) that can be used to further understand the processes of structure formation in the EoR. Intensity maps can also be used as a cosmological probe, since the fluctuations in the intensity of emission/absorption lines are correlated with the underlying dark matter density fluctuations. Prime-Cam will map the spatially-integrated signal on clustering scales at $z > 6$ (i.e. a few arcmin$^2$), which is done most efficiently with 6 meter class telescopes that have a spatial resolution matched to the key structures, along with a very wide FoV.

The spectroscopic-imaging module, EoR-Spec, inside Prime-Cam will probe the EoR via the cooling radiation in the 158 µm [CII] line from star-forming galaxies. This line is typically the brightest line in star forming galaxies [8, 9], directly maps the large-scale three-dimensional distribution and properties of the sources of reionization [10, 11], and is unaffected by IGM absorption or dust extinction, both of which hamper traditional UV/optical diagnostic tracers [12]. The Prime-Cam survey is designed to obtain a tomographic map of the EoR through the aggregate clustering signal in the power spectral density from faint star-forming galaxies from $z = 3.3$–9.3 (see Fig. 1). We will map a 16 deg$^2$ region in at least two deep multi-wavelength extragalactic surveys fields (e.g. E-COSMOS and CDF-S/H-UDF; see Figure 5 and Table 2). Backup fields are chosen across the full LST range to be observed under conditions when 350 µm observing (as required by the other surveys) is not possible, including daytime observing. The full survey area will likely be sufficient to obtain a significant detection of the clustering signal out to $z = 7$–8 (e.g., Refs. [13, 14, 15]; see Fig. 1), and to enable cross-correlation with HI 21 cm maps from experiments like HERA covering some of the same fields [15]. Such cross-correlations will allow us to further track the development of ionization bubble structures.

At the same time that we measure the [CII] line, EoR-Spec also covers at least two CO rotational lines in emission across virtually the entire $z$ range, and at least four lines at $z > 1$. The CO rotational ladder line luminosity typically peaks at J ~ 5 to 8 in forming galaxies but is 10 to 100 times weaker than the [CII] line. Detected CO line emission will therefore likely arise from intermediate redshift ($z = 0$ to 3) systems, where it is an important tracer for molecular gas. CO line emission from intermediate redshift galaxies will compete with [CII] from high redshift galaxies, but it can be easily distinguished from [CII] at most redshifts by the regular (115 GHz/(1+$z$)) spacing of CO rotational transitions. Our survey will also cover [OIII] 88 µm emission at $z > 6.7$, which could be even stronger than [CII] in low-metallicity galaxies, but is likely weaker than lower-$z$ [CII] at the same observing frequencies. The [OIII] line arises from gas ionized by massive O stars. We will investigate the potential of [OIII] through in-band cross-correlation with [CII] at the same redshift.

One of the most critical technical challenges for [CII] intensity-mapping studies is to extract the large-scale information in the presence of foreground line emission. This will be achieved through a three-fold approach. First, we will directly mask pixels that contain known low to intermediate-redshift galaxies that are expected to exhibit CO emission. This will be achieved by specifically targeting the extragalactic deep fields with the best, densest spectroscopic coverage over 7-20

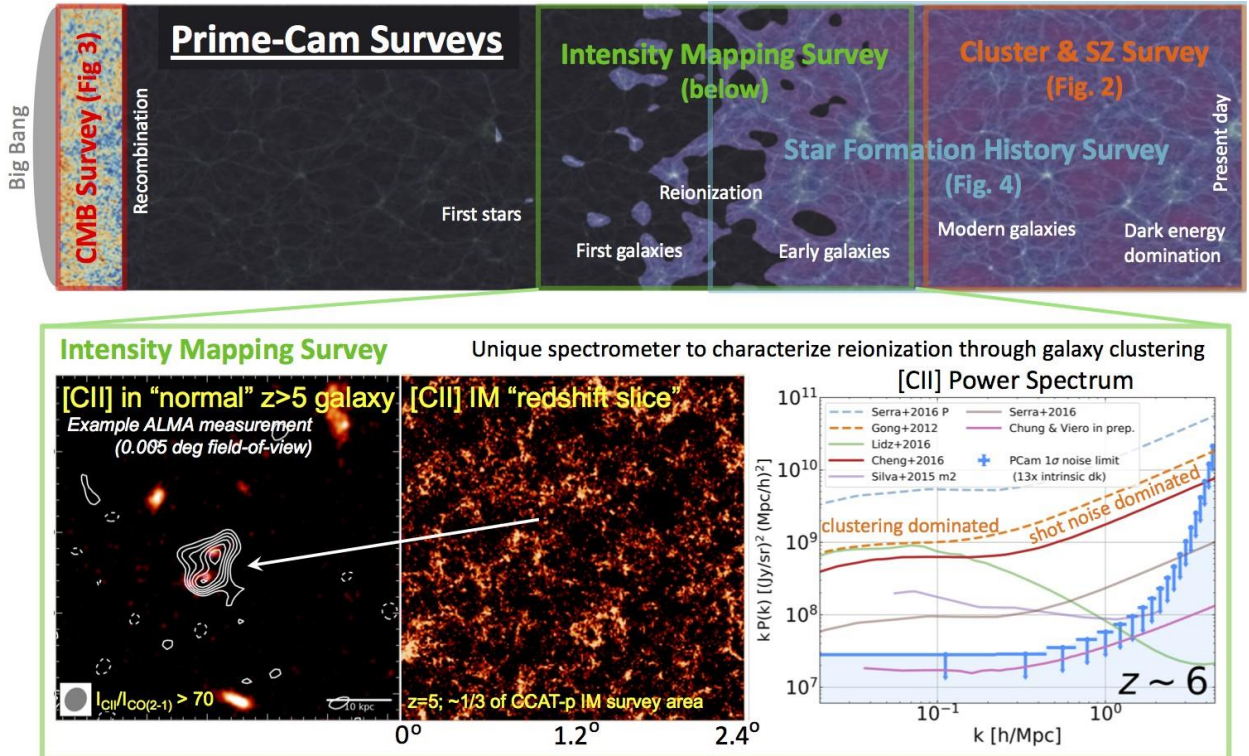

**Figure 1.** Prime-Cam measurement span an enormous range of cosmic times and scales, as shown on this schematic history of the Universe (adapted from [18,19]). *(top):* The four primary science surveys addressed by Prime-Cam observations include: intensity mapping of the epoch of reionization surveys (EoR; lower panels), galaxy cluster and Sunyaev-Zeldovich surveys (SZ; Fig. 2), CMB polarization surveys (CMB; Fig. 3), and dust-obscured star formation history surveys (SFH; Fig. 4). These surveys will complement many existing and upcoming observations (Fig. 5). *(bottom):* EoR survey, simulated "redshift slice" (*middle*) of the area covered by the [CII] intensity mapping survey in the EoR, representing a small spectral bin within the vast $z = 3.3–9.3$ redshift range covered by the EoR-Spec [CII] measurements. The brightest of the sources in some of the pixels are those now detected in [CII] on an individual basis with ALMA (*left*; [20]). Besides the bright sources, Prime-Cam will also detect the aggregate [CII] emission from the highly numerous but individually faint reionization galaxies that are below the detection threshold of even ALMA and JWST. The entire FoV of ALMA is approximately 4 × smaller than each of the 6,000 Prime-Cam pixels, and its bandwidth is more than 30 × smaller than what is covered by the EoR survey. Prime-Cam will measure the clustering of star forming galaxies at these redshifts (*right*), which determines the topology of cosmic reionization, through an analysis of the power spectrum of the [CII] emission (shown for $z=6.0$). State-of-the-art model predictions (lines) differ by factors of ~30–50 (Chung & Viero, in prep.; Refs. [17, 12, 21, 22, 13]). The full Prime-Cam survey will be sufficiently sensitive (blue curve of upper limits) to differentiate between these predictions for the first time.

deg$^2$ areas by the time of our survey (in particular, the new largely extended E-COSMOS and E-CDF-S regions; see Table 2). Based on models, we expect that this will require masking <10% of the total survey area [16]. Second, we will remove the contribution of CO foregrounds through in-band cross-correlation of frequencies where multiple CO lines are expected to appear. Third, we will be able to distinguish between [CII] and lower-$z$ CO emission through the anisotropic shape of their power spectra when projected onto a common coordinate frame, as shown in Ref. [17].

### 2.1.2 What is the nature of dark energy and the sum of neutrino masses?

The new frontier in cosmology is to utilize and extend the 6-parameter ΛCDM model to understand fundamental properties of the Universe, such as the nature of dark energy and the sum of neutrino masses. The redshift-independent Sunyaev-Zeldovich (SZ) effects, which can be efficient probes of structure growth, provide a compelling means to pursue these goals. Thermal SZ (tSZ) surveys have identified over a thousand galaxy clusters (e.g. [23] and references therein), enabling cosmological studies based on their number counts and angular correlations; however, current cosmological constraints are limited by systematic uncertainties in cluster properties. Prime-Cam measurements combined with 2–3 mm wavelength data from Advanced ACTPol [24] and Simons Observatory (SO, Ref. [25]) will find around 16,000 clusters and detect the fainter kinetic SZ (kSZ) effect to trace the peculiar velocity field in the Universe, a new complementary probe of dark energy and neutrinos [26, 27].

In addition, the unique submm coverage provided by CCAT-prime allows precise tracing of the SZ spectral shapes of individual clusters enabling measurements of their mean electron temperature (the relativistic SZ effect, rSZ). With these measurements CCAT-prime will provide valuable complementary information for deriving accurate cluster properties, like masses. Wide area Prime-Cam surveys (Table 1) will also provide independent and complementary constraints on dark energy and the sum of neutrino masses that are comparable to those of future experiments such as the Dark Energy Spectroscopic Instrument, DESI. The constraint on the sum of neutrinos masses is obtained by comparing the amplitude of fluctuations measured by low-$z$ surveys such as with Prime-Cam and DESI to that from the CMB at z = 1100, since neutrinos slow down structure formation in the low-$z$ Universe. Cosmological neutrino mass measurements will in general be limited by constraints on the optical depth since recombination [28, 29]. Despite this, there is a need to pursue multiple cosmological neutrino mass constraint approaches to probe beyond ΛCDM and potentially resolve the neutrino mass hierarchy problem [30].

The proposed Prime-Cam measurements will for the first time enable the detection of all three SZ components (yielding cluster optical depth, bulk velocity, and temperature) and dust emission from individual clusters [31, 32], for a large, statistically significant cluster sample. Among planned experiments, only CCAT-prime will provide observations with sufficiently broad coverage of the SZ spectrum (Fig. 2), and will do so with 5 to 20 times better resolution (depending on frequency) than Planck, the current standard for panchromatic SZ science [23]. In addition, spectral imaging of the brightest clusters near the zero-crossing of the thermal SZ spectrum at 220 GHz with EoR-Spec will enable SZ component separation with unprecedented precision, marking the evolution of this field towards using spectroscopic SZ science to understand clusters and cosmology.

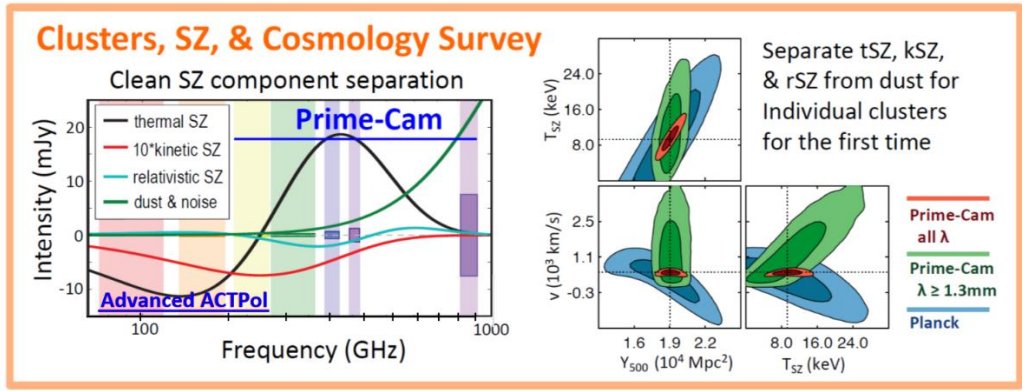

**Figure 2: SZ survey** – (*left*) For the galaxy cluster, SZ, and cosmology surveys, high frequency Prime-Cam bands above 230 GHz (left) are needed in addition to low frequency bands (which we will use from Advanced ACTPol and Simons Observatory) to remove both cluster and background galaxy dust emission to extract the temperatures and velocities of galaxy clusters (*right*, 1σ and 2σ contours shown) [31, 32].

Furthermore, Prime-Cam offers several opportunities for measurements of gravitational lensing, which probe dark energy and neutrinos through their effects on low-redshift gravitational potentials. Recent work indicates that lensing can potentially be measured in Prime-Cam's maps of both the CIB [34] and [CII] 158 µm line intensity [33] at several tens of standard deviations at least, constituting a first detection of lensing in each. Analysis methods from CMB lensing can straightforwardly be adapted for this goal [33, 34]. The resulting lensing maps can be cross-correlated with contemporary measurements from Advanced ACTPol and the Simons Observatory (SO), and also with clustering or lensing data from photometric surveys like LSST, to break internal degeneracies and isolate systematics. Prime-Cam is uniquely positioned to pioneer new detections of gravitational lensing, and open the door to harnessing their power as a cosmological tool.

### 2.1.3 How does feedback influence the formation and evolution of galaxies?

Tremendous advances have been made in our theoretical and numerical modeling of how galaxies form and evolve over cosmic time. These theoretical constructs are challenged and verified through observations of the physical and thermodynamical properties of the baryons in galaxies and clusters. High signal-to-noise cross-correlation measurements of the tSZ, kSZ, and rSZ effects provide independent windows into the thermodynamic profiles of individual galaxy clusters, ensemble averaged groups, and galaxies [35]. A key issue is quantifying and constraining the baryonic processes that make star formation globally inefficient, such as energetic feedback and non-thermal pressure support. Cross-correlations between CCAT-prime observations and galaxy, group, cluster, or quasar samples will probe these processes. Furthermore, these cross-correlation measurements will provide independent and essential tests for the successful cosmological hydrodynamical simulations that match the optical properties of large samples of galaxies from space and ground-based optical surveys like LSST.

The clean separation of the SZ effects and intrinsic thermal dust emission from galaxies and clusters in cross-correlation measurements requires multi-wavelength coverage from millimeter to submillimeter bands provided by CCAT-prime in order to avoid biased SZ measurements (e.g. [36]). Biases from extragalactic dust emission are some of the main systematic limitations for higher-order SZ statistics like the tSZ power spectrum, which provide additional information on the thermodynamic processes in galaxy formation, like feedback [37, 38]. Likewise, degeneracy with the cosmic dust emission also limits the accuracy in modeling the kSZ power spectrum (e.g. [39]), which is a highly promising probe to advance our understanding of the EoR epoch. CCAT-prime therefore opens a new unbiased observational window into the thermodynamic properties of galaxies and clusters all the way back to cosmic reionization.

### 2.1.4 What is the fundamental physics of the early Universe?

In recent years enormous progress was made using CMB temperature and polarization measurements to constrain cosmological parameters and characterize large scale structure. The CMB research community is developing plans for a next generation "Stage IV" CMB survey (CMB-S4) that might use the CCAT-prime and SO telescopes to achieve dramatic improvements in constraints on inflationary gravitational waves and neutrino properties [30]. CCAT-prime is a potential telescope platform for CMB-S4, and it also offers unique capabilities for important advances in high-frequency polarization science before CMB-S4. In particular, the excellent high-elevation site and telescope capabilities will enable sensitive measurements in the telluric windows between 0.1 to 1 THz. No other developed site or telescope on the Earth offers comparable sensitivity to CMB polarization and galaxy clusters spanning this frequency range, and no space or balloon telescopes offer comparable resolution.

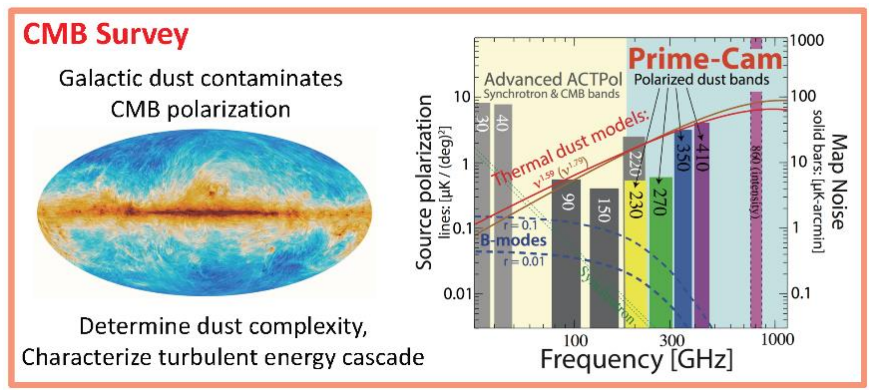

**Figure 3: CMB survey** – (*left*) CMB polarization measurements are contaminated by Galactic dust (like the Planck dust polarization map [40]) that plays a role in the turbulent energy cascade in molecular clouds in the Galaxy [43]. The polarization-sensitive detectors on Prime-Cam will be used to characterize the complexity of this dust to better understand our galaxy and, along with Advanced ACTPol and SO, to produce the cleanest CMB maps yet to constrain inflation and neutrinos [44]. The proposed Prime-Cam bands and sensitivity levels (shown for the pilot survey, Table 1) are complementary to Advanced ACTPol (*right*), which will enable rapid measurement advances by combining the data sets.

Prime-Cam will make the best measurements of CMB dust foregrounds at the higher frequencies required to understand the polarized Galactic dust that currently limit constraints on inflation from BICEP/Keck and Planck [40]. Recent work indicates that measurements above 250 GHz may be essential for the CMB lensing measurements needed to improve constraints on inflation [42]. In addition, low-resolution gravitational wave searches suffer from the fundamental limitation that a single beam on the sky will contain many small foreground structures with varying dust properties, such that the averaged emission from different regions will no longer be a power law, but will have a correction such as a running slope [48]. Averaging effects within beams like these and along the line of sight can be a serious challenge for low-resolution experiments [49, 50]. By measuring dust polarization properties above 250 GHz with sub-arcmin resolution and high signal-to-noise ratios using Prime-Cam, we will be able to predict the spectrum and polarization angles of dust emission averaged over $1°$ pixels and remove dust polarization from low-resolution inflationary gravitational wave experiments. The proposed survey will also improve our understanding of Galactic dust, magnetic fields, and the turbulent energy cascade that leads to star formation in the Galaxy [43]. Prime-Cam data combined with Advanced ACTPol and SO data will lead to the cleanest CMB maps yet, which are expected to provide unbiased measurements of CMB lensing to enable the delensing of the primary CMB anisotropies and constrain inflation [30].

Prime-Cam's access to very high frequencies opens up yet another unexplored window of physics from new scattering mechanisms. In addition to Thomson scattering, CMB photons also undergo Rayleigh scattering with neutral hydrogen and helium, which produces a strongly frequency-dependent cross-section and is a definite prediction of standard cosmology. Rayleigh scattering of CMB photons has the effect of a frequency-dependent change to the visibility function, suppressing temperature fluctuations and generating new polarization at high frequencies and small angular scales. Rayleigh scattering probes the early Universe by providing access to new modes [45] and improvements on cosmological parameters affecting recombination physics [46, 47]. Including Rayleigh scattering can improve $\Lambda$CDM parameter constraints by around 10% over primary CMB constraints. Prime-Cam, with its wide frequency range and survey area (Fig. 5), is uniquely positioned to make the first detection of Rayleigh scattering of the CMB and utilize the information from it in future cosmological constraints. If this approach is successful with Prime-Cam, it could enable substantial improvements in this top priority science goal for CMB-S4 [30]. CCAT-Prime's high frequency capabilities will be critical in mapping the Rayleigh signal and digging it out of the foregrounds.

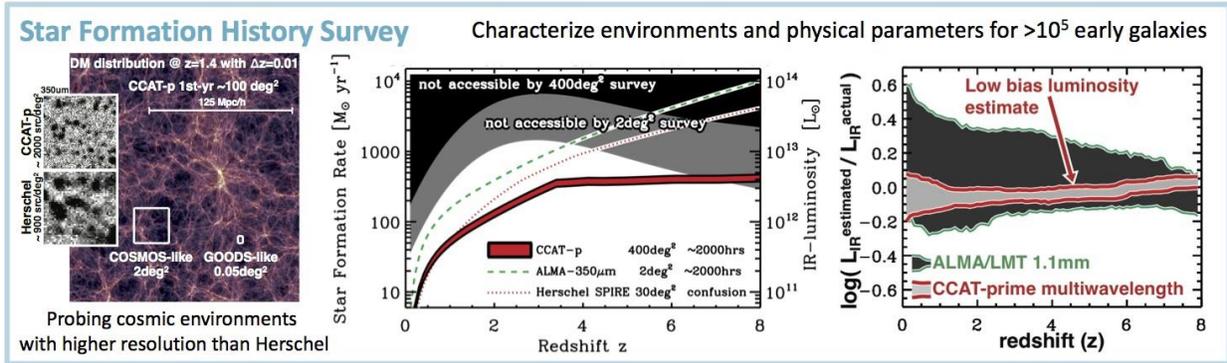

**Figure 4: SFH survey** *(left)* Wide area multi-wavelength CCAT-prime surveys will measure the dust-obscured star formation properties of hundreds of thousands of galaxies out to $z \sim 5$ to $7$ with high precision. This will greatly enhance our picture of cosmic galaxy evolution, probing 2 to 10 times lower luminosities than *Herschel*-SPIRE at $z > 2$ (due to source confusion; see insets superposed on Millennnium simulation [58]) while sampling large-scale environments (voids, the field, groups, and clusters) over two orders of magnitude larger areas than feasible to map with ALMA (the FoV of ALMA is a fraction of a single CCAT-prime pixel). *(middle)* Simulated CCAT-prime survey luminosity limits, compared to surveys feasible with existing instrumentation [59, 60, 61, 62]. CCAT-prime probes 2 to 10 times deeper than *Herschel*-SPIRE while sampling environments and luminosities inaccessible to ALMA surveys because of cosmic variance (gray areas). *(right)* Predicted reliability of far-infrared luminosity measurements, highlighting the vast improvements with CCAT-prime's 350 $\mu$m band and wide spectral coverage.

### 2.1.5 How do galaxies form over cosmic time?

Stars form from the collapse of molecular clouds. Much of this star formation is hidden from our view due to dust within the clouds that absorbs starlight, heats up, and reradiates the power at far-infrared wavelengths. Half of the starlight emitted through cosmic time is obscured by dust, which we measure as the cosmic infrared background (CIB; [51, 52]). Therefore, to understand the star-formation history of the Universe, one must measure it both directly through the optical emission from stars (which is redshifted into the near-infrared bands), and indirectly from the dust (which is redshifted into the submillimeter bands). About 80% of the CIB from star formation emitted at redshifts ≤ 1.7 was resolved into individual galaxies detected in at least one of the bands observed by Spitzer and Herschel [60]. Yet at earlier times, where measurements are limited by the selection function of short wavelength (i.e., < 250 µm) observations, and the relatively poor sensitivity of current long wavelength observations, the fraction resolved drops to just ~10% [60, 53].

CCAT-prime will reach 2 to 10 times deeper than the Herschel surveys (which are limited by confusion noise), securely detecting hundreds of thousands of galaxies at redshifts potentially as high as 7 (Fig. 4). CCAT-prime will unveil about 40% of the CIB at 350 µm, while being dominated by high-redshift (z > 1) galaxies [53]. Its wide-field (> 400 deg$^2$) surveys sample environments on large scales that cannot be mapped with ALMA [54], and reach down to luminosities that remained inaccessible to Herschel (Fig. 4). The 350 µm band is critical to extract physical properties (in particular infrared luminosities and dust-obscured star formation rates) to far greater precision than possible with long-wavelength surveys alone (e.g., the Large Millimeter Telescope, LMT; Fig. 4).

Combining our surveys with synergistic work in the optical/near-infrared (LSST, DES, Euclid and WFIRST) we will be able to identify key parameters that regulate star formation (such as environment and matter content) over time. We expect to directly reveal the mergers of gas-rich galaxies that triggered violent and short-lived starbursts which may have created the local giant elliptical galaxies [55, 56] and in a statistical manner reveal the star formation processes leading to "normal" galaxies like the Milky Way in the epoch of their assembly ($z \approx 5$). In addition, the wide-field survey promises to uncover the most intense starbursts in the Universe, including those entirely missed by even the deepest optical/near-infrared surveys [57]. The proposed survey will cover the rich, multi-wavelength GAMA fields (around 100 deg$^2$ total) in the pilot phase to the confusion limit at 350 µm (Fig. 5). "Parallel" ~4 to 16 deg$^2$ continuum images obtained with the EoR intensity mapping fields will reach far below the confusion limit, providing a precise measurement of the confusion noise on the scales most relevant to CCAT-prime. The full survey will also cover the extended SDSS Stripe-82 area (previously covered only shallowly by Herschel/SPIRE) to comparable depths. In total, we expect to detect more than 600,000 individual galaxies above a signal-to-noise ratio of 5 in the full survey, 1,200 of which we expect to be at redshifts 5–8.

| Table 2: Overview of planned extra-galactic survey regions (Figure 5) and observing parameters | | | | | | |
|---|---|---|---|---|---|---|
| Survey | Field ID | LST range | Area (deg$^2$) | Time (hr) | Sensitivity (at representative $\nu_{obs}$(GHz)) | Supporting Surveys[d] |
| EoR[a] | E-COSMOS | 7.0-13.0 | 7.2 | 1600 | 0.02 MJy sr$^{-1}$ bin$^{-1}$ at 220 | 1 |
|  | E-CDFS | 23.5-7.5 | 10 | 2400 | 0.02 MJy sr$^{-1}$ bin$^{-1}$ at 220 | 2 |
|  | HERA-Dark | (filler) | 10 | (backup) | 0.02 MJy s$^{-1}$ bin$^{-1}$ at 220 | 3 |
| SFH[b] | GAMA09 | 5.5-12.5 | 36 | 120 | 2.5 mJy beam$^{-1}$ at 860 | 4 |
|  | GAMA12 | 8.5-15.5 | 36 | 120 | 2.5 mJy beam$^{-1}$ at 860 | 4 |
|  | GAMA15 | 11.0-18.0 | 36 | 120 | 2.5 mJy beam$^{-1}$ at 860 | 4 |
|  | Stripe 82 | 20.0-5.5 | 300 | 1000 | 2.5 mJy beam$^{-1}$ at 860 | 5 |
| SZ/CMB[c] | AdvACTPol | All | 12000 | 4000 | 20 µK√s (CMB units) at 270 | 6 |

LST ranges are given for elevations of >40°. See Table 3 for more sensitivity details. [a]Spectroscopy; sensitivities provided assuming R=300. [b]Continuum; will also serve as deep calibration fields for SZ/CMB survey. [c]Continuum; will also be used to select the rarest, brightest objects for SFH survey. [d]Data sets: (1) Deep Subaru HSC+PSF spectroscopy (in progress) and COSMOS X-Ray-to-meter-wave multi-wavelength survey; (2) deep Euclid grism spectroscopy (upcoming), HERA HI 21 cm (upcoming), & H-UDF/CDF-S multi-wavelength surveys (incl. JWST GTO); (3) HERA HI 21 cm (upcoming), VLASS; (4) GAMA, H-ATLAS Herschel/SPIRE, ACT, VLASS; (5) SDSS, HeLMS/HeRS Herschel/SPIRE, VLASS; and (6) ACT/Simons (in progress/upcoming), Planck, SDSS, DES, DESI (upcoming), LSST (upcoming).

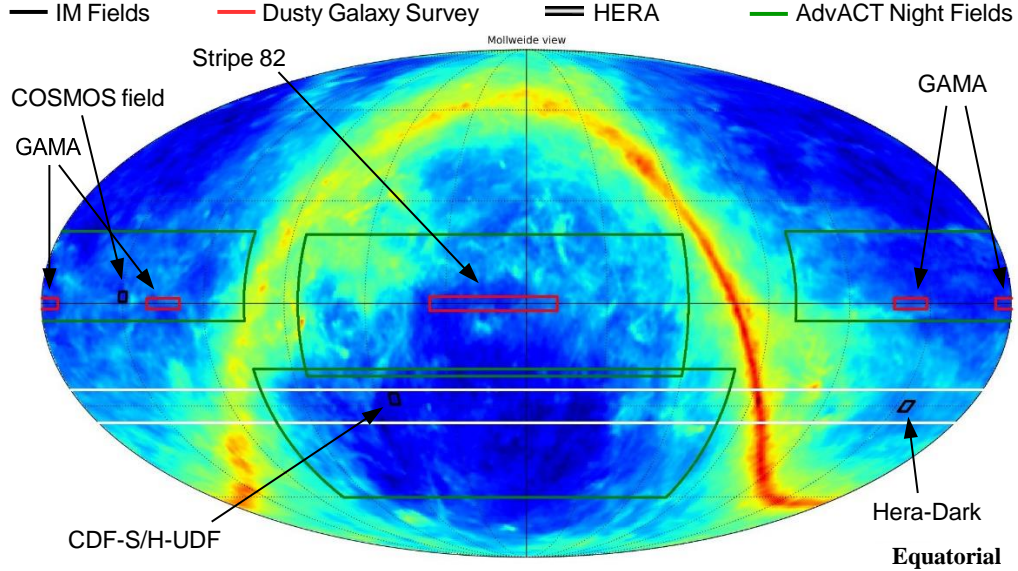

**Figure 5: Overview of relevant survey fields.** Preferred survey fields are indicated for the galaxy clusters and cosmology (*green boxes*), star formation/dusty galaxy (*red*), and intensity mapping/reionization (black, gray, *small white*) surveys (see Table 2 for survey areas, LST ranges, and integration times). The full HERA field (white) is shown for reference, within which we pick a region with minimal background emission, as indicated by the Planck dust map ([63]; logarithmic color scale). Days will be shared between the dusty galaxy and reionization surveys to optimize weather usage and LST coverage. Full days will be spent on the cosmology fields. Survey sessions will be scheduled to allow for approximately equal exposure times between surveys over the course of a year, assuming that no daytime observing is possible at 350 $\mu$m, and that 50% of the night time weather is appropriate for observations below 850 $\mu$m.

## 2.2 Science with CHAI and Prime-Cam: Star Formation in the Milky Way and Nearby Galaxies

### 2.2.1 Molecular Gas Reservoirs: CO Dark Gas.

Since stars form within molecular clouds to understand star formation one needs to understand the relationship between molecular gas mass and star formation efficiency. For nearly 50 years, the workhorse tracers of molecular cloud mass have been the low-J lines of CO [64]. Even so, it was realized 30 years ago that CO is likely a very poor tracer of molecular gas mass in low metallicity systems [65]. This is because molecular clouds are often illuminated by stellar UV radiation which will photodissociate CO at the surface, forming $C^0$ and $C^+$. The penetration of these dissociating and ionizing photons is determined by dust absorption, whereas the $H_2$ molecule is self-shielding against the dissociating radiation. Therefore, assuming the dust to gas ratio tracks the gas phase metal abundances, the CO dissociation region will be driven deeper into lower metallicity clouds resulting in a CO emitting core that is much smaller than the size of the $H_2$ molecular cloud. The first observations of the [CII] line emission from the low metallicity Large Magellanic Cloud (LMC) showed a surprisingly large [CII]/CO line intensity ratio confirming this effect [8]. Very large [CII]/CO ratios have since been shown to be a common property of low metallicity systems [66]. More recent models show that during the formation of a molecular cloud from atomic gas, CO takes much longer to form than $H_2$ [67] so that when clouds are young, much of the $H_2$ will not be traced by CO: a chemical timescale cause for the prevalence of what has come to be known as CO-dark gas [68]. With the advent of the Herschel spectrometers, multiple line tracers of molecular clouds became available [69] which showed that for the Galaxy, about half the total interstellar medium may reside in this CO-dark gas [70].

A promising tracer appears to be atomic carbon which, while classically associated with stronger PDR regions, has also been observed toward cold, dark clouds [71, 72], showing that it is very widely distributed. This is supported by detailed simulations ([73], Fig. 6). Thus, atomic carbon, since it can be mapped with CHAI on CCAT-prime on very wide scales (most of the Galactic Plane, and Magellanic Clouds, Table 2) with good sensitivity and velocity resolution, appears to be a very good proxy for at least a significant fraction of the CO-dark gas. Hence it allows us to characterize the distribution, excitation and velocity structure of this ubiquitous ISM component. In particular, this will give insight into cloud formation processes, for example Ref. [72] observe a significant discrepancy between the atomic carbon and $C^{18}O$ distributions in their map), and also shed light on the distribution of cosmic rays, since they affect the C/CO ratio [74].

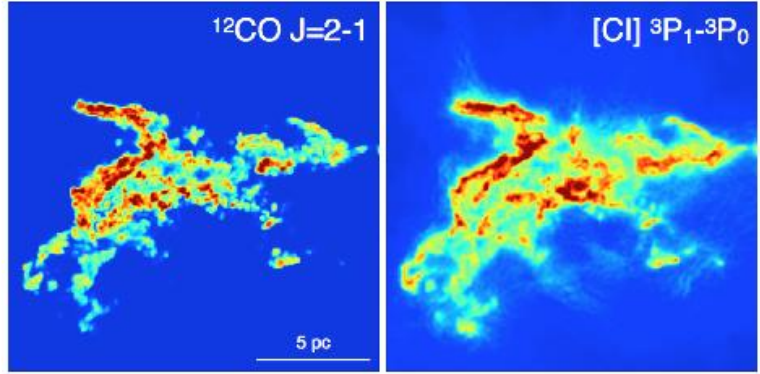

**Figure 6:** Simulation of the CO J = 2-1 (*left*) and [CI] $^3P_1$ - $^3P_0$ (*right*) integrated intensity for a GMC exposed to an average interstellar ultra-violet radiation field in the inner Galaxy [65]. In these models, the [CI] traces the cloud material in both the high density inner regions traced by CO, as well as the extended, diffuse regions that are free of CO emission but still have significant $H_2$ abundance. Simulations courtesy of F. Molina, and S.C.O. Glover.

CCAT-prime can map much of the Galactic plane (200 deg.$^2$) and the Magellanic Clouds (90 deg.$^2$) in the [CI] (1-0) line at sensitivities appropriate for characterizing the CO-dark gas. These surveys will be accompanied by smaller zoom-ins in [CI] (2-1) to determine the excitation, and with maps of mid-J CO(4-3) of the same area, and smaller maps of higher-J (e.g. 7-6) CO lines in the nearby Gould Belt clouds. The higher-J lines can be shock-excited and therefore characterize the dissipation of turbulence in molecular clouds.

The pair of [CI] lines have great potential to trace the bulk of the molecular gas in nearby galaxies. Over the last decade the Herschel Space Observatory made a significant contribution to the number of existing extragalactic regions observed in atomic carbon (e.g., the "Beyond the Peak" survey [75]). However, the nature of the molecular gas traced by [CI] in these kilo-parsec scale extragalactic regions is still unclear. For example, from the analysis of a large sample of luminous infrared galaxies find that the [CI](1-0) line is as a good as a tracer of the cold molecular gas as CO(1-0) [76]. In contrast, a study of luminous infrared galaxies and starburst nuclei indicates the bulk of the [CI] line emission arises from a dense gas component that represents only a fraction of the total molecular gas mass in the galaxy [77]. Future observations of the [CI] lines with CHAI on CCAT-prime can target extragalactic regions for which we have robust measurements of the total molecular gas mass from the combination of dust continuum emission, low-J CO lines, and the [CII] 158 μm transition. These will be key to assess the reliability of the [CI] lines as molecular gas tracer, determine the physical conditions of the molecular gas emitting the [CI] lines, and quantify variations of the [CI] lines emission as a function of galactic environment and metallicity. All these measurements will be of great help to the interpretation of the growing number of observations of the molecular gas in high-z galaxies using the [CI] and CO transitions.

### 2.2.2 Variability/Transient Studies with Prime-Cam

With the advent of large-format, extremely sensitive, submm detectors, it is now possible to efficiently monitor astronomical sources for brightness variations. The first dedicated submm survey, the JCMT Transient Survey [78], has uncovered a half dozen protostellar variables ranging from quasi-periodic [79] through sources with a general brightening or dimming across years [80, 81]. Although limited by small number statistics, the survey results suggest that 10 percent of deeply embedded protostars vary in submm brightness by greater than 5 percent per year. Since the bulk of the luminosity being emitted by protostars and reprocessed by their enshrouding envelopes is due to mass accretion from the circumstellar disk onto the star, their variability is expected to directly trace episodic mass accretion events [82] and instabilities within the protoplanetary accretion disk (e.g. [83, 84, 85]). That significant brightness changes are being uncovered on multiyear timescales suggests that the inner several AU regions of these accretion disks are responsible for the variability [77], precisely the disk locations where interest in planet formation peaks.

Prime-Cam provides a powerful tool for variability and transient studies of the submm sky. The square degree FoV matches well the size of nearby star-forming clouds allowing for efficient snapshot imaging and precise relative flux calibration across epochs (e.g. [86]). Additionally, by operating at a much shorter wavelength, 350 μm, the variability signal due to episodic accretion will be significantly enhanced over the JCMT survey. CCAT-prime will more closely monitor variation in the source brightness due to accretion $L_{acc}(t)$ whereas the present 850 micron measurements monitor the temperature response of the envelope, more weakly scaling as $L_{acc}^{1/4}$ ([82], Yoo et al. in preparation). Finally, the enhanced sensitivity of Prime-Cam will allow for significantly fainter sources to be monitored, including protostars at later stages of evolution where much of the envelope has dissipated. Thus, a multi-year CCAT-prime survey of protostars will quantify the importance of episodic accretion, and its underlying timescales, throughout the protostellar phase. Regular monitoring of more distant Galactic star-forming regions should provide a large enough sample (e.g. [87]) to uncover more rare, but extremely energetic events such as FUors and Exors [88, 89] which can then be followed up with JWST and ALMA.

Finally, CCAT-prime may also be sensitive enough to detect rare extragalactic transient events provided the observing strategy and source extraction software are carefully optimized. Detection expectations for tidal disruption of sub-stellar objects around supermassive black holes have been calculated for the LMT and ALMA [90] and need to be refined for CCAT-prime. Millimeter-wave discoveries of outbursts from AGN (e.g. [91]) and GRBs (e.g. [92]) may be detectable and have been searched for by other survey telescopes (e.g. [93]).

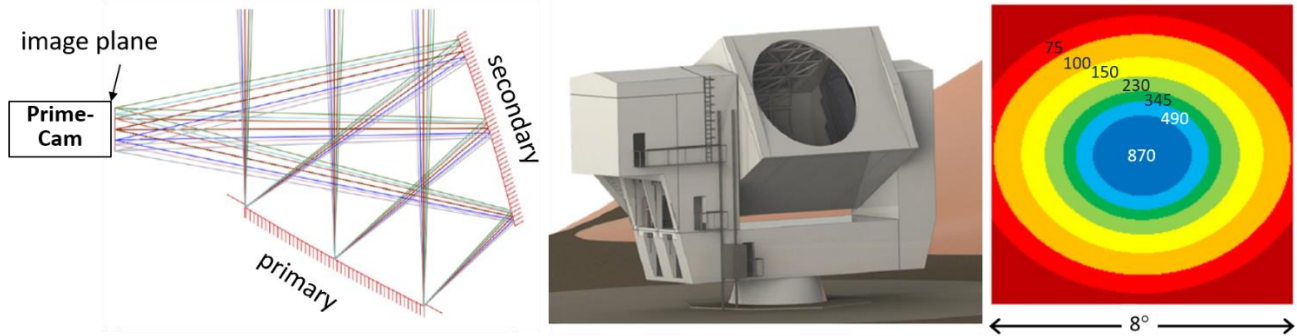

**Figure 7:** (*left*) Crossed-Dragone optical path from the sky (top) to the image plan (left) and Prime-Cam instrument. (*center*) rendering of the CCAT-prime telescope at the 5600m elevation site. The yoke altitude-azimuth structure is 23×8×16 meters in size and is expected to weigh about 220 metric tons. (*right*) Diffraction-limited FoV of the CCAT-prime image plane as a function of frequency. The colors show the areas within which the Strehl ratio is >80% at 870, 490, 345, 230, 150, and 75 GHz from blue to red respectively.

## 3. THE CCAT-PRIME TELESCOPE, OPTICS AND SITE

### 3.1.1 Telescope and Optical Design

The CCAT-Prime telescope optics are based on the crossed-Dragone concept first proposed in Dragone [2]. The off-axis optical system both enables a totally unobstructed view of the sky which minimizes telescope emissivity and uncontrolled sidelobes, and also leads to very low cross-polarization which makes it an excellent choice for CMB experiments. As such, the crossed-Dragone design was chosen for both the QUIET [94] and Atacama B-Mode Search experiments [95], and proposed for "Stage 4" CMB experiments [96]. Figure 7 shows the basic optical design. Light from the sky enters at the top and is reflected by the primary mirror to the secondary and imaged to a very large, flat focal plane. In the classic crossed-Dragone design, the primary and secondary are designed so that first-order aberration and cross-polarization caused by the tilt of the paraboloid figure of the tilted primary is canceled by the hyperbolic figure of the tilted secondary. For CCAT-prime we applied the Ritchey-Chrétien design that uses higher order polynomials in the figure of both mirrors leading to a remarkable improvement in the diffraction limited FoV – as much as a factor of 10 at 350 µm ([97], Fig. 7).

The telescope primary and secondary mirrors have clear apertures of 6.0 and 5.7 m, and are segmented into 87 and 78 square panels respectively, each with surface areas near $0.5m^2$. Panels are machined from a single block and light-weighted to about 5 kg each, and supported by a truss structure. The telescope housing and drives are robust structures to enable the rapid scanning on the sky necessary to minimize sky noise and achieve near background-limited performance from the cameras and spectrometers. See Parshley et al. [97] and [98] for more details on the optics and telescope design.

### 3.1.2 The site: Why go to 5600 m elevation?

The CCAT-prime site has an elevation of 5600 m, which is 600 m above the ALMA array site up the steep sides of the Cerro Chajnantor mountain. This presents many logistical challenges (that we are, or expect, to be meeting). Why go to 5600 m? Success in submm astronomy from the ground is all about the telluric transmission above, which is dominated by the rotational absorption lines of water. The ALMA site on the Chajnantor plateau is an excellent site for submm astronomy as demonstrated from the science returns from, for example, the Atacama Pathfinder Telescope (APEX, [99]) and the ALMA array itself [100]. However, a balloon-borne radiosonde campaign in the late 1990's showed that the exponential scale height for precipitable water vapor (pwv) for balloons launch on the plateau at 5000 m elevation is 1130 m leading to the expectation that much could be gained by moving to the > 5600 m elevation summits of nearby peaks [101]. Subsequent decades-long simultaneous tipper radiometer measurements at the CCAT-prime site near the peak of Cerro Chajnantor, the plateau, Mauna Kea, and South Pole demonstrated that the Cerro Chajnantor site is truly exceptional, with the best top two weather quartiles of all in zenith opacity in the short submm 350 μm band ([1], [102]). Converting 350 μm zenith opacity to zenith pwv, the Cerro Chajnantor site has only about half that of the plateau (and South Pole) in the first quartile, and 60% of the plateau in the second quartile (Fig. 8) which improves transmission (at 45° source elevation) by factors of 1.5 and 2.0 in the $1^{st}$ and $2^{nd}$ weather quartiles respectively. In the mm-wave bands, the effect is significantly less dramatic since the transmission (at 1 mm, for example) is always above 87% and 92% for the top two quartiles respectively. However, since the emissivity of the sky, $\varepsilon_{sky} = 1-\eta_{sky}$, even the subtle change from 92% (plateau) to 95% (Cerro Chajnantor) can have significant effects on sensitivity system.

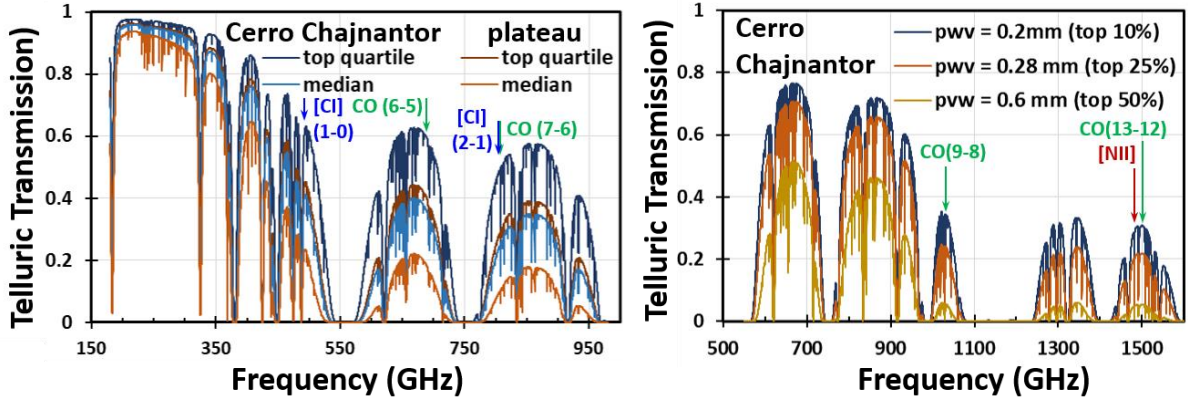

**Figure 8:** *(left)* Telluric transmission comparison between the CCAT-prime site at 5600 meters elevation on Cerro Chajnantor and on the Atacama plateau at 5000 m over the frequency range addressed with Prime-Cam and CHAI obtained with the APEX calculator (http://www.apex-telescope.org/sites/chajnantor/atmosphere/transpwv/). *(right)* CCAT-prime site in the best water vapor conditions showing the available THz windows. A few astrophysically important transitions are shown on both plots.

For a properly designed camera or spectrometer, working with background-limited detectors, the system sensitivity, often called noise equivalent flux, or NEF ($W/m^2$-$s^{1/2}$) is a function of the system emissivity, $\varepsilon_{system}$ which includes the sky emissivity (= 1- sky transmission), the telescope emissivity, and the dewar window emissivity. In most cases this reduces to NEF $\propto \varepsilon_{system}^{1/2}$ (for $h\nu \gg kT$), or NEF $\propto \varepsilon_{system}^{1}$ (for $h\nu \ll kT$). The latter is the case for Prime-Cam. Therefore, if the sky emissivity is small, say a few percent, then the telescope and window emissivity become important. This is the case at mm wavelengths for CCAT-prime, so that it is critical to keep these emissivities very small. We have worked to keep our telescope and dewar window emissivity < 2%. These excellent telescope/instrument parameters mean that even in the mm bands, CCAT-prime benefits from the modest increase in sky transparency (from 97 to 99%) gained from going from the plateau to the Cerro Chajnantor site. Clearly the sensitivity gains in the short submm bands will be much greater still. Figure 9 shows the expected gains in mapping speed (proportional to $1/NEF^2$) for CCAT-prime at the Cerro Chajnantor site compared with the plateau site

## 4. CCAT-PRIME INSTRUMENTATION

### 4.1.1 First light instrumentation

We have two instruments planed for first light: Prime-Cam and CHAI. Prime-Cam is a large FoV camera and imaging spectrometer based on direct detection bolometer arrays, and CHAI is a multi-channel dual color high resolution heterodyne spectrometer. For details of Prime-Cam see Vavagiakis et al. [3].

### 4.1.2 Prime-Cam

Prime-Cam is a direct detection system that contains up to 7 optics tubes, each of which can be independently targeted for a particular science case (Fig. 9). These optics tubes will be enclosed in a single cryostat cooled with a combination of pulse-tubes that provide the 80 K, 40 K, and 4 K cold sinks necessary for cooling optics and reducing internal radiation and heat loads, and a dilution refrigerator that supplies the 100 mk cold sink necessary to operate the sensitive bolometer arrays. The central tube will contain the shortest wavelength optics tube the 860 GHz (350 μm) camera thereby taking advantage of the high Strehl ratios in the center of the CCAT-prime image plane necessary for sensitive, diffraction-limited operation . Arranged in a ring around the central tube are the 6 other optical tube positions. These will likely be filled with a combination of camera tubes appropriate for the SZ, SFH, and CMB surveys, SZ/CMB-Pol and the imaging Fabry-Perot interferometer (FPI), EoR-Spec that appropriate for intensity mapping in the EoR and spectrophotometry of cluster SZ.

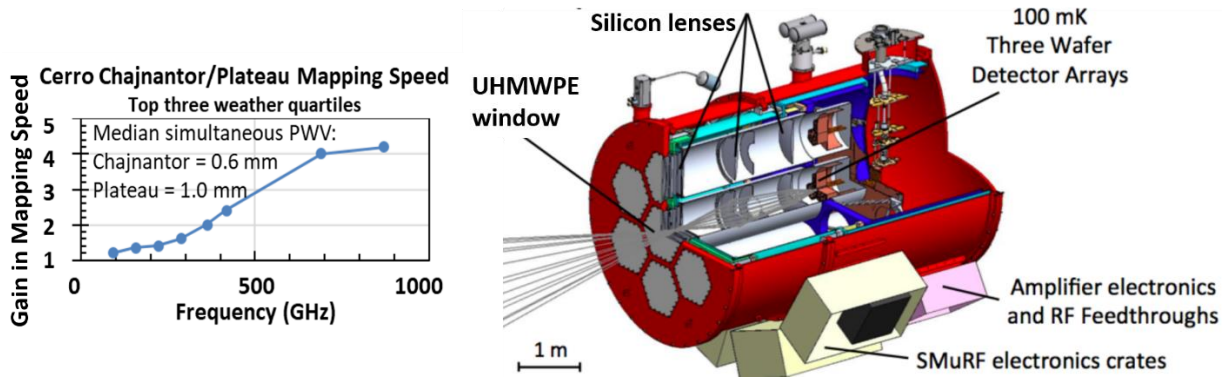

**Figure 9:** *(left)* Gains in mapping speed realized by going to 5600 meters elevation on Cerro Chajnantor over the same instrument on the Atacama Plateau *(right)* Prime-Cam concept. The entrance windows will be anti-reflection coated (ARC) UHMWPE, cold lenses are high purity silicon with metamaterial ARC [107]. The dewar is approximately 1.8 m in diameter and 2.8 m long.

**Star Formation History (SFH) Camera:** The primary science case for the 860 GHz SFH camera is spatially resolving dusty star forming galaxies at high redshifts. Therefore, spatial resolution is premium and we require dense packing of pixels. We also need many pixels to make wide fields on the sky, so at 860 GHz, we will use kinetic inductor detectors (KIDs) whose simplicity enables about a factor of three times closer packing of pixels than TES arrays. In this way we can fabricate about 6000 pixels spaced by 1.5 f/#·λ on each of three -150 mm diameter silicon wafers for a total pixel count over 18,000 within this optics tube. The availability of multi-thousand pixel close-packed arrays together with recent advances in array performance consistent with background limited performance also motivate this choice [103,104,105].

**SZ/CMB-Pol:** For the SZ/CMB-Pol camera and EoR-Spec we have selected quad-chroic bolometers which promise background limited performance, and have pixel counts per 150-mm silicon wafer sufficient to deliver our science [106, 24]. These bolometers are feedhorn fed and coupled into a superconducting integrated circuit through an ortho-mode transducer. In this way, each spatial pixel delivers power to 8 independent TESes (4 colors and 2 polarizations) greatly increasing mapping speeds [108]. Dichroic versions of these detectors have been successfully deployed on ACT-Pol [24]. For SFH science the two polarizations at each frequency will be added together for higher sensitivity.

**EoR-Spec:** EoR-Spec will use an array of quad-chroic bolometers very similar to that of the SZ/CBM-Pol camera, but since polarization is not of interest, the dual polarization outputs will be tied to the same TES to reduce the TES-count. The goal of EoR-Spec is to cover the frequency range from 190 to 450 GHz as efficiently as possible over as wide a FoV as possible. The EoR signal is quite faint and must be mapped over a 16 deg$^2$ region, so many (> 10,000) detectors will be required for a ground-based experiment to reach the required sensitivity levels in 4000 hours observing time. This pixel count can be distributed in large instantaneous bandwidth spectrometers with relatively low spatial positions [109], or with relatively modest instantaneous bandwidth spectrometers with large numbers of spatial positions on the sky. A key consideration for extremely faint line detection over broad regions of the sky are the systematics that will eventually overwhelm photon noise. It is arguable that the best system will have a good cross-product of spectral-spatial multiplexing so as to minimize systematics over the two spatial and one spectral dimension.

Our EoR-spec concept is based on an imaging FPI and achieves very good spatial *and* spectral multiplexing. The mechanical design is based on a flex-vane translation stage (Fig. 10, Refs. [110,111,112,113]), and we plan on using silicon substrate based mirrors as reflectors [114]. We chose to use the quad-chroic bolometers so that we can filter the resonant orders of the FPI with the narrow-pass bolometer bands and simultaneously image at the four frequencies selected by the bolometers. We set the FPI with a fundamental resonant mode at 63 GHz. It will also resonant at integer multiples of this frequency so that transmission fringes appear at 189 GHz (3rd order), 252 GHz (4th order), 315 GHz (5th order) and 378

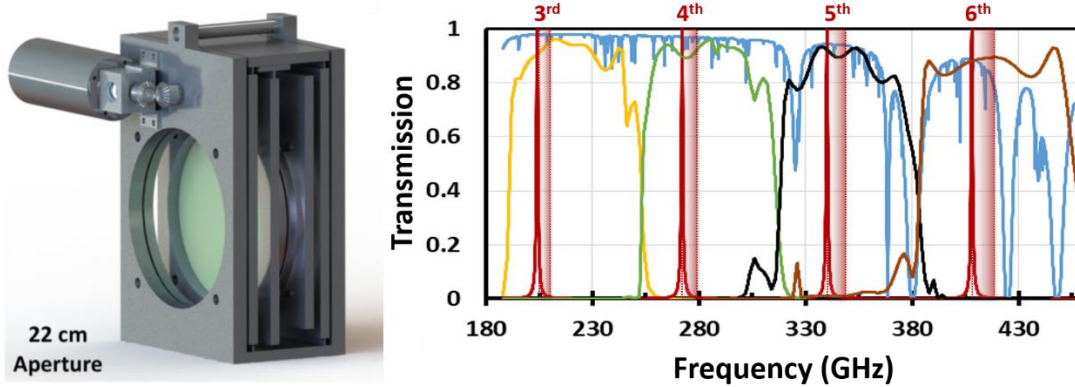

**Figure 10:** (*left*) mechanical drawing of flex-vane based FPI suitable for use in the EoR-Spec. (*right*) Four (n = 3, 4, 5, 6) FPI resonant orders (red Lorentzian profiles) superposed on the (yellow, green, black, and red) colored transmission profiles of the quad-chroic detectors. The bandpasses are well matched to the telluric transmission spectrum (background blue trace). The instantaneous bandpass of the FPI due to tilts of the field in the pupil plane are indicated as transparent rectangles.

GHz (6th order) (Fig. 10). Rays going through a FPI etalon at different angles will see a different resonant frequency (to the blue) than the on-axis rays. Therefore, at any region of the focal plane that is off the optical axis the resonant frequency is slightly different. This effect is mitigated by putting the FPI at a pupil where all the rays from a single point on the sky are collimated and go through the FPI parallel. There is still, of course, some wavelength shift within each diffraction limited beam, since such a beam subtends a finite angle on the sky. This will limit the resolving power that one can obtain. By conservation of eténdue, these angles are made smaller by the ratio of the collimated beam diameter to the primary mirror diameter. For a collimated beam size of 20 cm, this effect limits the resolving power in a beam to < 1000 which is safely larger than our required resolving power of 280.

The finite collimated beam size also results in an advantage with respect to spectral multiplexing. Beams that are off the optical axis will see center frequencies shifted with respect to the central beam. For a 20 cm collimated beam this amounts to shifts totaling 33 GHz in the focal plane for the 4 orders of the array, or about 13% of the total 255 GHz bandwidth we wish to sample. Since our resolving power is ~ 280, we instantaneously sample 38 spectral resolution elements on the sky, and only need 10 fixed settings of the FPI to address the entire 255 GHz bandwidth required for our science.

### 4.1.3 CHAI

The CCAT-prime Heterodyne Array Instrument, (CHAI, Fig. 11) will be a two-color heterodyne array receiver. It covers the part of the 650 µm atmospheric window containing the carbon monoxide (CO) $J = 4\rightarrow3$ line and the neutral atomic carbon fine-structure line [CI] $^3P_1\rightarrow{}^3P_0$. In the 370 µm window CO $J = 7\rightarrow6$ and [CI] $^3P_2\rightarrow{}^3P_1$ can be observed simultaneously. With 64 pixels in a square array, the instrument will cover a field of view of 7.5 × 7.5 arcmin$^2$ and of 4.5 × 4.5 arcmin$^2$ at the two wavelengths, respectively. The receiver footprint is overlaid on a [CII] map of the Horsehead nebula [115] in Figure 11.

CHAI occupies one of two auxiliary receiver spaces below the optical axis of the telescope. To access the focal plane, it uses a partly retractable optics setup that can be removed for prime-Cam observations. All-reflective optics direct the light into the receiver cabin, where it is split by polarization to serve the two wavelength bands. Each color has its own independent cryostat with pulse-tube coolers. A third cryostat location is foreseen to allow temporary operation of alternative instruments.

CHAI uses on-chip balanced SIS-mixers [116] combined in 4-pixel mixer blocks. The LO signal is distributed by waveguides. The IF output band covers 4-8 GHz. It is amplified and directly coupled to digital Fast-Fourier-Transform Spectrometers (FFTSs) [117].

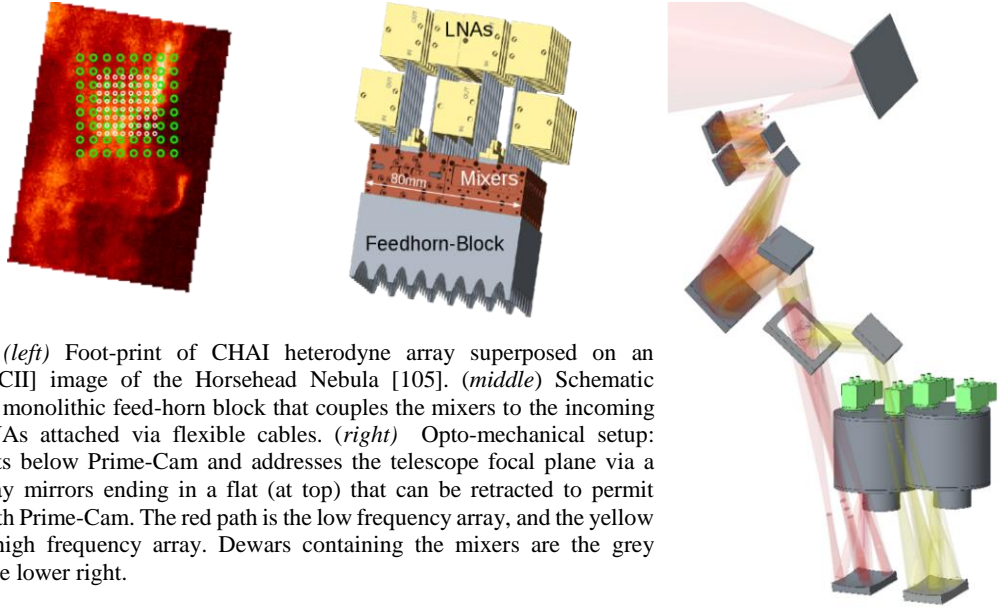

**Figure 11:** *(left)* Foot-print of CHAI heterodyne array superposed on an upGREAT [CII] image of the Horsehead Nebula [105]. *(middle)* Schematic drawing of a monolithic feed-horn block that couples the mixers to the incoming radiation LNAs attached via flexible cables. *(right)* Opto-mechanical setup: CHAI mounts below Prime-Cam and addresses the telescope focal plane via a series of relay mirrors ending in a flat (at top) that can be retracted to permit observing with Prime-Cam. The red path is the low frequency array, and the yellow path is the high frequency array. Dewars containing the mixers are the grey cylinder at the lower right.

Table 3 contains the expected per beam sensitivities for Prime-Cam on CCAT-prime. We evaluate for the top two weather quartiles for a source at 45° elevation and include all losses. Table 4 lists system parameters for CHAI.

| Table 3: Prime-Cam sensitivities Top 2 weather quartiles[1] Elevation = 45° | | | | | | |
|---|---|---|---|---|---|---|
| System Properties | | | Sensitivity | | | |
| Band (GHz) | Beam (") | l.o.s trans. 1st (2nd)[2] | NEFD[3] (mJy-s$^{1/2}$) | NET/beam[4] (mK-s$^{1/2}$) Ray-Jeans | CMB[5] | NEI (MJy/sr-s$^{1/2}$)[3] |
| 861 | 14 | 44% (23%) | 152 (420) | 1.2 (3.4) | | |
| 405 | 37 | 71% (59%) | 70 (110) | 0.55 (0.87) | 13 (21) | 4.8 (7.8) |
| 350 | 39 | 87% (79%) | 32 (47) | 0.29 (0.43) | 3.6 (5.3) | 2.9 (4.3) |
| 270 | 46 | 95% (92%) | 13 (16) | 0.14 (0.18) | 0.72 (0.92) | 1.6 (2.0) |
| 220 | 53 | 96% (94%) | 10 (12) | 0.12 (0.23) | 0.37 (0.46) | 1.0 (1.2) |

Notes: [1]CCAT-prime quartiles – zenith pwv: 1st = 0.28 mm, 2nd = 0.60 mm, 3rd = 1.0 mm; [2]First values, 1st quartile, second value (in parenthesis) refer to 2nd quartile pwv; [3]Dual polarization detection; [4]Single polarization detection; [5]$\delta T_{CMB} = \delta T_{RJ} \cdot ((e^x-1)^2/(x^2 e^x))$, where $x = h\nu/(kT_{CMB})$.

| Table 4: CHAI Heterodyne Array Receiver System Parameters | | |
|---|---|---|
| Parameter | Low Freq. Array | High Freq. Array |
| RF range [GHz] | 455-495 | 800-820 |
| Noise temp (DSB) [K] | <100 | <200 |
| IF band [GHz] | 4 – 8 | 4 – 8 |
| Resolution [kHz]/[km/s] | 100/0.06 | 100/0.04 |
| Velocity coverage (km/sec) | 2500 | 1500 |
| Beam size ["] | 26 | 15 |
| Array Format (pixels) | 8 × 8 | 8 × 8 |
| Field of view [arc. Min. | 7.5 × 7.5 | 4.5 × 4.5 |

### 4.1.4 Second Generation Instruments

It is expected that CCAT-prime will have at least a 20 year lifetime, so that there is will be several generations of instrumentation used on the facility consistently becoming more powerful as technologies evolve. The modular optical design for our first light instrument means that within itself, Prime-Cam has great scientific agility. Any system that fits within its 40 cm diameter, 1.4 m length tubes and is adequately serviced by its thermal sinks from 80 K to 100 mK is a possible option. Future options for Prime-Cam upgrades and for full 7.8° diameter FoV systems include:

1. Bringing Prime-Cam to its full potential through the addition of 4 optics tubes to join the first light triplet. We anticipate that all our current science cases will continue to be impactful when the first surveys are complete and motivate further deeper and wider-field work. A likely upgrade fleshing out the entire 7 optics tube configuration is a total of 3 SZ/CMB-Pol tubes, 3 EoR-Spec tubes and a SFH camera.

2. Expanding the CHAI footprint to 128 × 128 pixels, and the frequency coverage to lower frequency windows.

3. Exploiting the site for work in the 1.5 THz (200 μm) windows. A 200 μm *camera* would be uniquely powerful for exploration of dust emission in the Milky Way and nearby galaxies as 200 μm approaches the dust emission peak so it places constraints on dust temperature and obscured star formation, or protostellar, luminosity.

4. Opening the THz windows for spectroscopy with a 0.5° diameter imaging FPI with resolving power of 2000, or with a heterodyne array. The [NII] 205 μm line can be readily mapped in the Milky Way and nearby galaxies. This line in of itself is used as a proxy for Lyman continuum photons for low-density ionization bounded HII regions [118]; with [CII] it discriminates the fraction of the [CII] line emission that arises from ionized gas [119]; and with the [NII] 122 μm line, it constrains low-excitation ionized gas densities and mass [119, 120]. The high-J CO lines (13-12 and 12-11) are also detectable in both Galactic star formation regions where they trace shocked, often outflowing gas, and in ULIRG or AGN nuclei where they are signposts of an X-ray dominated region (XDR) in a warm, dense molecular tori that may envelope, and ultimately feed the engine [121].

5. Installing superconducting resonator based broad-band direct-detection spectrometers [e.g. 122, 123]. These "spectrometers on a chip" have small footprints so that in principle thousands of full bandwidth spectrometers can be employed for line intensity mapping of high redshift galaxies. If such spectrometers come to fruition, they provide an efficient method of pursuing intensity mapping science with different systematics than EoR-Spec.

6. Using the outer 1.5° annulus of focal plane not engaged by Prime-Cam by installing 12 new optics tubes, nearly tripling the FoV to the full 7.8° allowed by CCAT-prime's optics. The new, very large cryostat required could use the 7 tubes from Prime-Cam. Populating the total of 19 tubes at their diffraction limited wavelength of operation with 1.5 f/#·λ pixels would bring CCAT-prime's total spatial pixel count to > 270,000 for single color pixels, or about 570,000 if we use quad-chroic devices at the lower frequencies. Prime-Cam will already have major impact on CMB-S4 surveys with its high frequency measurements. The new (near) mega-pixel system would have powerful low frequency capabilities which would be particularly useful for the cluster SZ and CMB science outlined above.

## 5. CCAT-PRIME TIMELINE

The construction and installation of CCAT-prime at the 5600 m site on Cerro Chajnantor is fully committed with the signing of contractual details with Vertex Antennentechnik in July 2017. The project is currently between our preliminary design review (PDR) which was passed in April 2018, and the critical design review (CDR) that is scheduled for October 2018, with a final design review (FDR) expected in March 2019. We are working with the Tokyo Astronomical Observatory (TAO) to complete the road to the summit in 2018, and expect the concrete pad for the telescope to be installed in the spring of 2019. The fabrication period is 15 months long, which partially overlaps an 8 months trial assembly by Vertex Antennetechnik in Germany. By October 2020 we expect the first telescope parts to begin arriving at the site, for a 5 month assembly and checkout leading to our first light in 2021.

**Acknowledgements:** CCAT-prime funding has been provided by Cornell University, the Fred M. Young Jr. Charitable Fund, the German Research Foundation (DFG) through grant number INST 216/733-1 FUGG, the Univ. of Cologne, the Univ. of Bonn, and the Canadian Atacama Telescope Consortium. NASA grant NNX16AC72G supports FPI development at Cornell.